\documentclass[10pt, conference]{IEEEtran}

\usepackage{amsmath}
\usepackage{subfigure}
\usepackage{amssymb} % mathfrak
\usepackage{enumerate}
\usepackage{graphicx}
\usepackage{tikz,pgf}
\usepackage{pgfplots, xstring} % for decoration
\usepackage{pgfplotstable}
\usepackage{soul}
\usetikzlibrary{calc}

\def\Kin{K_{in}} % population
\def\P{\mbox{Pop} } % population

\def\a{`\textit{a}' }
\def\ab{`\textit{a}'}        %used when \a precedes a punctuation mark DSB
\def\MAll{M}     % used to be M_a(A)
\newcommand{\mydot}{.}
\newcommand{\eq}{=}

\begin{document}
\onecolumn
\title{Dissortative From the Outside, Assortative From the Inside: Social Structure and Behavior in the Industrial Trade Network}

% alternative title:
% the role of missing information in network creation

\author{
\IEEEauthorblockN{Guy Kelman}
\IEEEauthorblockA{Racah Institute for Physics\\
Hebrew University of Jerusalem\\
Jerusalem, Israel\\
Email: superk@cs.huji.ac.il}\\   %<------ Line breaks in the current column
\IEEEauthorblockN{David S. Br\'{e}e}
\IEEEauthorblockA{Department of Computer Science\\
University of Manchester\\
Manchester, UK\\
Email: davidSbree@gmail.com}
\and
\IEEEauthorblockN{Eran Manes}
\IEEEauthorblockA{Department of Public Policy\\
Guilford Glazer Faculty of\\Business and management\\
Ben Gurion University\\
Be'er Sheva, Israel\\
Email: msemanes@gmail.com}\\   %<------ Line breaks in the current column
\IEEEauthorblockN{Marco Lamieri}
\IEEEauthorblockA{Research dept.\\
Intesa SanPaolo\\
Milan, Italy\\
Email: marco.lamieri@intessanpaolo.com}
\and
\IEEEauthorblockN{Natasa Golo}
\IEEEauthorblockA{Racah Institute for Physics\\
Hebrew University of Jerusalem\\
Jerusalem, Israel\\
Email: natasa.golo@gmail.com}\\   %<------ Line breaks in the current column
\IEEEauthorblockN{Sorin Solomon}
\IEEEauthorblockA{Racah Institute for Physics\\
Hebrew University of Jerusalem\\
Jerusalem, Israel\\
Email: sorin@huji.ac.il}
}

\maketitle


\IEEEpeerreviewmaketitle

\begin{abstract}
  It is generally accepted that neighboring nodes in financial networks are negatively assorted
  with respect to the correlation between their degrees. This feature would play an
%  Financial networks are generally considered negatively assorted
%  in respect of the degree-degree correlations. This fact plays an
  important `damping' role in the market during downturns (periods
  of distress) since this connectivity pattern between
  firms lowers the chances of auto-amplifying (the propagation of) distress.
  In this paper we explore a trade-network of industrial firms where
  the nodes are suppliers or buyers, and the links are those invoices that the
suppliers
  send out to their buyers and then go on to present to their bank for discounting.
  The network was collected by a large Italian bank in 2007, from their
  intermediation of the sales on credit made by their clients.  The network also shows dissortative
  behavior as seen in other studies on financial networks.  However,
  when looking at the \textit{credit rating} of the firms, an important attribute internal to each
  node, we find that  firms that trade with one another share overwhelming similarity.

%  On top of the sampling bias that is mainly due to the fact that
%  the bank targets specific firms,??? 

  We know that much data is missing from our data set.
  However, we can quantify the amount of missing data using \textit{information exposure}, a variable that connects
  social structure and behavior.
  This variable is a ratio of 
  the sales invoices that  a supplier presents to their bank
  over their total sales.

  Results reveal a non-trivial and robust relationship between
  the \textit{information exposure} and \textit{credit rating} of a firm,
  indicating the influence of the neighbors on a firm's rating.
%  Medium credit rated firms expose less information about their
%  buyers, while both extreme credit-worthy or credit-constrained firms
%  expose full information about their buyers.
  This methodology provides a new insight into how to reconstruct a network
  suffering from  incomplete information.

%  missing at random
\end{abstract}

\section{Introduction}

%  When one has a snapshot in time of a system, since
  The topology of a network is the visible result of integrative processes
  in the underlying system. It may be possible to deduce the dynamics of the underlying system 
  from such a network. However, the mechanism
%  when investigating beyond the individual link the network
  so deduced will be extremely sensitive to
  any data that is missing from the network. If the information on the system is incomplete %, even at random,
  the rendered network may provide a misleading picture
  of the system and impact our understanding of the mechanisms that
  created it.

   Financial networks are known for being negatively assorted,
   i.e. neighbouring nodes in the network are dissimilar, in
   particular as regards the  degree of their in- and out-links.
   Among practitioners and economists this property is desired because
   it renders the financial network robust to percolation (propagation
   of distress or growth). The knowledge that contagion rarely
   happens may catch us by surprise when financial shocks
   do indeed propagate from the local level to the national/international
   level. In the events preceding the 2008 financial crisis, small
   systemic shocks affected large proportions of the industrial and
   financial networks. The usual response of firms to market downturns
   was then amplified and the response swept across the network using
   the monetary (communication) channels. One reason for our lack of
   control over this incident was that a proportion of the communication
   channels that link  peers were not known to the banking
   system: the high risk mortgages were traded in the market but
   the credit-unworthy clients behind them remained anonymous.

%   \hl{We wanted to deduce the propagation of distress experienced by some firms to others} \textcolor{red}{DO WE?}.
  We use a network of asymmetric links to
   represent the exchange of goods/services for financial payment
   between firms (nodes); it is similar to the better known
   communication networks. % , and the source for asymmetry roots in the roles of each node (buyer or seller).
   Each link is between a supplier firm and one of its customers who bought a product or service from them.
%   These data were collected by a bank,
   Our data contain a snapshot in time of many firms that provide
   goods and services in exchange for financial payments in the
   year 2007. This was the year when financial crises occurred
   global-wide which led to an economic downturn.

   These data were collected by a single large Italian bank,
   fulling its function as an intermediating agent in a delayed
   payment procedure. The bank recorded the names of the two parties
   and the amounts of money that one firm, the buyer, owes the
   other, the supplier.

   Apart from the bank, both the supplier and the buyer will have
   entered the face value of this contract into their own bookkeeping systems
   under the items \textit{accounts receivable} and \textit{accounts
   payable}.

%  Since the links are  asymmetrical, firms appear asymmetrically on file. The bank prefers to deal with suppliers rather than buyers, with large firms, rather than small, and with the credit-worthy, rather than the credit-constrained. Thus already there is a selection bias on the part of the bank.
  The bank prefers to deal with suppliers rather than buyers, with large firms, rather than small, and with the credit-worthy, rather than the credit-constrained. Thus already there is a selection bias on the part of the bank.

  There is, moreover, another source of bias to the information on the
  network related to a social element. The
  reader could appreciate that a firm's choice of trading partner should be strategical.
  First and foremost because firms buy goods that they need. Second,
  our data describe the outcome of a short-term relationship between suppliers and buyers: the credit, or delayed payments, mechanism.
  Suppliers would prefer to work with
  buyers that clear their debt on time. However, financially constrained
  suppliers will probably be willing to work with less reputable
  clients in order to keep their business active.

  In later sections we will see that the result of combining these
  biases leaves a rather empty picture of the full network structure.
  Others working on financial trade networks have made attempts to
  predict the missing links. Ohnishi et. al. \cite{Ohnishi:2009uq},
  for example, adapt a trade network by filling in at random so
  they could run standard flow solvers. This kind of missingness
  that we observe should be handled carefully, and link prediction
  should probably be avoided.

  Before we did the analysis we expected to find \textbf{(1)} that there
  would be long production chains; \textbf{(2)} that firms would present
  \ul{all} their bills of sale for discounting; so \textbf{(3)} there would
  be no difference in the ratio of bills presented/total sales
  between firms with different credit ratings. And finally,
  \textbf{(4)} that the credit ratings of suppliers and buyers would
  \ul{not} be correlated.

%  The intuition for these draws from theory in managerial economics.
%  Specifically, the principal (or process) of profit maximization.
  
%  When we combine the degree distribution (that shows similarity to social networks) and internal node attributes like credit-rating, we find that pairing is strategical.
%  \textcolor{red}{MOVE THIS, IT'S A RESULT} \hl{The data are also biased by an attribute that is jointly exhibited by pairs of firms. The invoices that are accepted by the bank as collateral for loans to the suppliers not only exhibit the internal attributes discussed above, but also a very important joint attribute: The credit rating of the buyer matches with the credit rating of the supplier. Pairs with non-matching credit-rating scores are rare.}
%  When we combine the degree distribution (that shows similarity to social networks) and internal node attributes like credit-rating, we find that pairing is strategical.
%  \textcolor{red}{MOVE THIS, IT'S A RESULT} \hl{The data are also biased by an attribute that is jointly exhibited by pairs of firms. The invoices that are accepted by the bank as collateral for loans to the suppliers not only exhibit the internal attributes discussed above, but also a very important joint attribute: The credit rating of the buyer matches with the credit rating of the supplier. Pairs with non-matching credit-rating scores are rare.}

  \subsection{Dissortative networks}
    Assortative mixing in networks is a term describing the correlation
    of `popularity' between different nodes. Popularity is a
    feature attributed
    to a node and measured by the number of incoming links to it.
    A network is positively assorted if the number of incoming links to
    a node is positively correlated with the number of incoming links to
    its neighbors. In assorted networks
    messages can spread within a small number of steps
    since there are many redundant
    links via which a message could travel.
    Negatively assorted networks contain highly connected nodes
    that are positioned sparsely throughout the network. Thus, in this type of topology the fast
    spread of messages is less likely \cite{Newman:2002fk}.
    However, if attributes of nodes are known, it
    is possible to combine structural and behavioral information
    for efficient routing inside this network \cite{Simsek:2008kx}.
    
  \subsection{The trade-credit network} 
  In our data, the network contains a record of financial interactions between peers.
  The interaction under investigation is recorded when a \textit{discount} process occurs. The bare explanation of a \textit{discount} on an invoice is that an owner of an invoice will sell it to a financial institute for a lower price than its face value. The buyer of the invoice will be the new creditor and will take upon himself the risk that the debtor will become insolvent.  This risk is combined into the rate of discount.

  Today, banks offer their customers a cheaper alternative to selling their
  trade bills. A customer of the bank can `collateralize its accounts
  receivable': the bank, instead of buying the invoice, 
  will extend a loan. The collateral for this loan will be the face value of the invoice.
  The customer of the bank is a firm that approaches the bank for
  a loan. Usually, when using discounted invoices, this firm will be a
  supplier, not a buyer. Extensive reviews of the reasons why
  this may be so were suggested by Omiccioli \cite{Omiccioli:zr}
  and Marotta \cite{Marotta:2000ys}.
  Here we note one obvious reason: the supplier needs to
  secure funds only for production of the goods/services, whereas the buyer
  needs to cover the total amount of the invoice. An invoice amount
  includes not only the costs of production but also the supplier's
  profit, a larger amount than production only. Thus, the amount
  by which the bank discounts is, for the supplier, part of his
  profits, but for the buyer it is part of his costs. Lower amounts
  on loan impose less risk on the lender and in return a more affordable 
  discount is offered.
  %financing a buyer is riskier to the bank.

%  \hl{trade-credit data base, balance sheets, 345,000 buyers, 140,000 suppliers also buyers - appears in} \ref{sec:Data}

  \subsection{Credit rating and financial costs}
  In order to facilitate an efficient discounting mechanism the
  banks created a credit-rating procedure. When the customer of the bank requires
  a loan, he should qualify as credit-worthy, i.e. a borrower that
  is financially capable of paying back. Credit-rating is a score provided to all banks by an external entity\footnote{CeBi - Centrale dei Bilanchi, a financial analyses service for the Italian banking system.}; it is usually between 1 and 9: 1 describes the most credit-worthy and 9
  describes the most credit-constrained borrower. This parameter is estimated
  every year from balance sheet items of the firm, and thus it is
  specific to the firm. 

  Credit-rating affects the terms on loans.
  When a firm believes that its bank is imposing conditions that
  are unreasonable, it may resort to other means of financing.
  In general, a firm's bank is the sole provider of loan financing and
  by declining a loan based on its terms (interest rate)
  the firm must consider other financing channels, the most intuitive
  of which is trade-credit; the firm will ask to delay its
  debt to suppliers and collect immediate payments from its buyers
  \cite{Petersen:1997kx}.
  By doing so a social component is added to the pool of financing channels, and this can be traced on the network of trading firms.

  The total interest paid in one year appears in the balance sheet
  of the firm as `Financial Costs'. \textit{Financial Costs}
  normalized by \textit{Total Bank Loans} corresponds with
  credit-rating. We will use this as a continuous proxy to the
  credit-rating score. The in-degree, being the number of buyers for each supplier, will be the social component,
  and derivations for quality of a firm's neighborhood will be combinations
  of degree and attributes internal to the nodes and the links (such as the
  sums of money that move between firms).

  \subsection{Missing data - not at random}

  In order to understand this next section we need to make a clear distinction
  between path length and the length of a supply chain. The path length
  between two nodes `o' and `d' is a network measure that counts
  the number of nodes needed to pass in order to reach node `d'
  from node `o'. Some nodes may be unreachable if they reside in
  different connected components or if the network is directional, and
  they reside in different strongly-linked components.

  A \textit{production chain} (also called a \textit{supply chain})
  is a directional chain of inputs. Goods pass in the `downstream'
  direction and money in the `upstream' direction.
  In general, a firm takes input from various suppliers and generates
  a single output product so the structure of a supply chain is
  like a tree subgraph.
  It is usually drawn from the wide end on the bottom (raw materials) to
  the root, or roots, on the top. The root level contains the firms
  that manufacture the end product(s). Each level of the tree
  corresponds to several intermediate inputs and is termed in network
  language `rank'.
  
  Since our data set is an account of the flow of money, different
  accounting procedures along the supply chain can cause the visible
  part of the chain to break apart, leaving smaller number of levels
  on each tree subgraph.

  A computerized algorithm that traverses the paths will
  disregard a broken supply chain since the nodes reside in
  different strongly-connected components. It is expected that the
  algorithm will uncover longer paths between the root of the supply
  chain and the end product.
  The average length of a supply chain was estimated
  by %[??] to be 4.3
 % (\hl{REFERENCE PRODUCTION CHAIN - this is a PhD work of someone,
 % and I cannot find it. I did find this
  Nair and Vidal {\cite{Nair:2011fk}} to be between 3.0 to 4.4.

  %\hl{We know this since the path lengths    in the network are much shorter than the typically observed lengths of production chains}. \textcolor{red}{need to explain that the path length IS shorter than the production chain. Average path length in the network is still lot longer than expected. Maybe put down THE LENGTH OF DIRECTIONAL CHAINS} % 

    Since the network and the consequential dynamics that we wish
    to deduce are both extremely sensitive to missing link information,
    and  since extending the network at random is likely to be a biased, given the bias already in the data due to the financial support procedure of
    the bank, we need to explore another
    method for uncovering the structure of the underlying trade
    network \cite{Huisman:2009fk}. Rhodes and Jones \cite{Rhodes:2008uq}
    provide an interesting method for generating links using
    probabilistic inference (Bayesian); however, they impose strong
    assumptions on the priors (e.g. the proportion of links in the
    system that are known). Others, like Smith and Moody
    \cite{Smith:2013kx}, point out that generating missing information
    out of a known but incomplete sample is a hard task since the
    different statistical measures behave with varying magnitudes
    across network topologies (i.e.  if we were but to know what
    the topology really is, then a probabilistic method would help
    greatly to improve the sample).

%  Information is missing from the inter-firm trade network.  \hl{We know this since the path lengths    in the network are much shorter than the typically observed lengths of production chains}. \textcolor{red}{need to explain that the path length IS shorter than the production chain. Average path length in the network is still lot longer than expected. Maybe put down THE LENGTH OF DIRECTIONAL CHAINS} % (\hl{REFERENCE PRODUCTION CHAIN - this is a PhD work of someone, and I cannot find it})
%    Since the network and the consequential dynamics that we wish to deduce are extremely sensitive to missing link information, and  since extending the network at random, given the bias already in the data due to the financial support procedure of the bank, is likely to be a biased, we need to explore another method for uncovering the structure of the underlying trade network \cite{Huisman:2009fk}. Rhodes and Jones \cite{Rhodes:2008uq} provide an interesting method for generating links using probabilistic inference (Bayesian); however, they impose strong assumptions on the priors (e.g. the proportion of links in the system that are known). Others, such as  Smith and Moody \cite{Smith:2013kx}, point out that generating missing information out of a known but incomplete sample is a hard task since the different statistical measures behave with varying magnitudes across network topologies (i.e.  if we were but to know what the topology really is, then a probabilistic method would help greatly to improve the sample).
%The above para cut as it is a repeat of the previous para, DSB.
    Firm size is suspected to be endogenous to the bias in the
    sample. The reason is that the bank may try to attract large
    firms rather than small ones. We can expect that large firms
    will deal more contracts (have more buyers), have larger production facilities
    (requiring larger total bank loans) and greater financial costs. In order
    to control for firm size we normalized quantities by
    the firm's total bank loans or by its size, using 
    the annual \textit{net-sales} (other common options that we might have used are total
    salaries and number of employees).

\section{Methods}
  \subsection{The data}
   \label{sec:Data}
   The two data sets at the bank which we could consult were :

   \begin{itemize}
     \item Time series of individual firm balance sheets (and Profit
     \& Loss statements) in the 8 years between 2002 and 2009.
     These data contain information that allows one to know the financial
     status of a firm. It will hereafter be abbreviated BS.
     \item Bank-mediated credit transactions of trading partners
     in the year 2007. 
     Each record contains 3 fields of interest:  the names the two transacting firms ($c, s$) and the total face
     value of all the trade-bills (invoices) passed between each supplier-buyer pair ($R_{cs}$), 
     as will be defined in section \ref{sec:definitions}.
     This data set will be termed TC.
%     This information is directed. i.e. any unique pair of firms
%     $A$ and $B$ may appear in two rows. Once when firm $A$ is
%     supplier and $B$ is buyer, and another time when $B$ is the
%     supplier and $A$ is buyer.
   \end{itemize}

   \noindent
   The data are not publicly available; they were directly accessible
   to only one researcher, who worked for the bank.  Programs to
   extract summary data from these two data sources were written
   by us and executed on a computer which belonged to the bank.
   The summary data, which we possess, were then further analysed
   to obtain the results reported here.

%  These data are proprietary and secret in part. So they were not directly
%  accessible to the research group. Most of the number-crunching
%  was done on the bank's computers. Only summary statistics and
%  aggregate plots were exposed outside of the bank.

   In total, in BS there are balance sheets of 1.3 million firms
   over the 8 year-window. On average, balance sheets of 700,000
   firms appear in any one year with overlap of approximately 300,000
   along the timeline. In 2007 there were 703,858 firms with
   net-sales greater than zero (potential suppliers) and 601,535
   firms had purchases greater than zero (potential buyers).
   
   In the TC data set there are 1,578,812 firms, connected by
   7,290,072 links.  When intersected with the firms in BS we obtain
   a total of 345,403 firms connected by 2,874,830 links. 273,726
   of the firms in the TC data are suppliers (have incoming links). And 
   of the joined data set TC+BS, 140,580 are suppliers. If we remove
   the suppliers that are linked to buyers without BS data, we are
   left with 129,584 suppler firms, all of whom have at least one
   buyer with BS information. We call this set $\MAll$.
   122,728 of the suppliers in $\MAll$ (94\%) have outgoing links and therefore
   are also buyers.  The remaining 215,819 firms are buyers only.

   An individual firm (node) assumes attributes from the BS such as
   firm size, credit-rating, financial costs or industrial classification.
   In the appendix we list the variables used as attributes.

\begin{figure}[h]
  \centering{
   \includegraphics[scale=0.2, clip, bb=20 10 1060 1043]{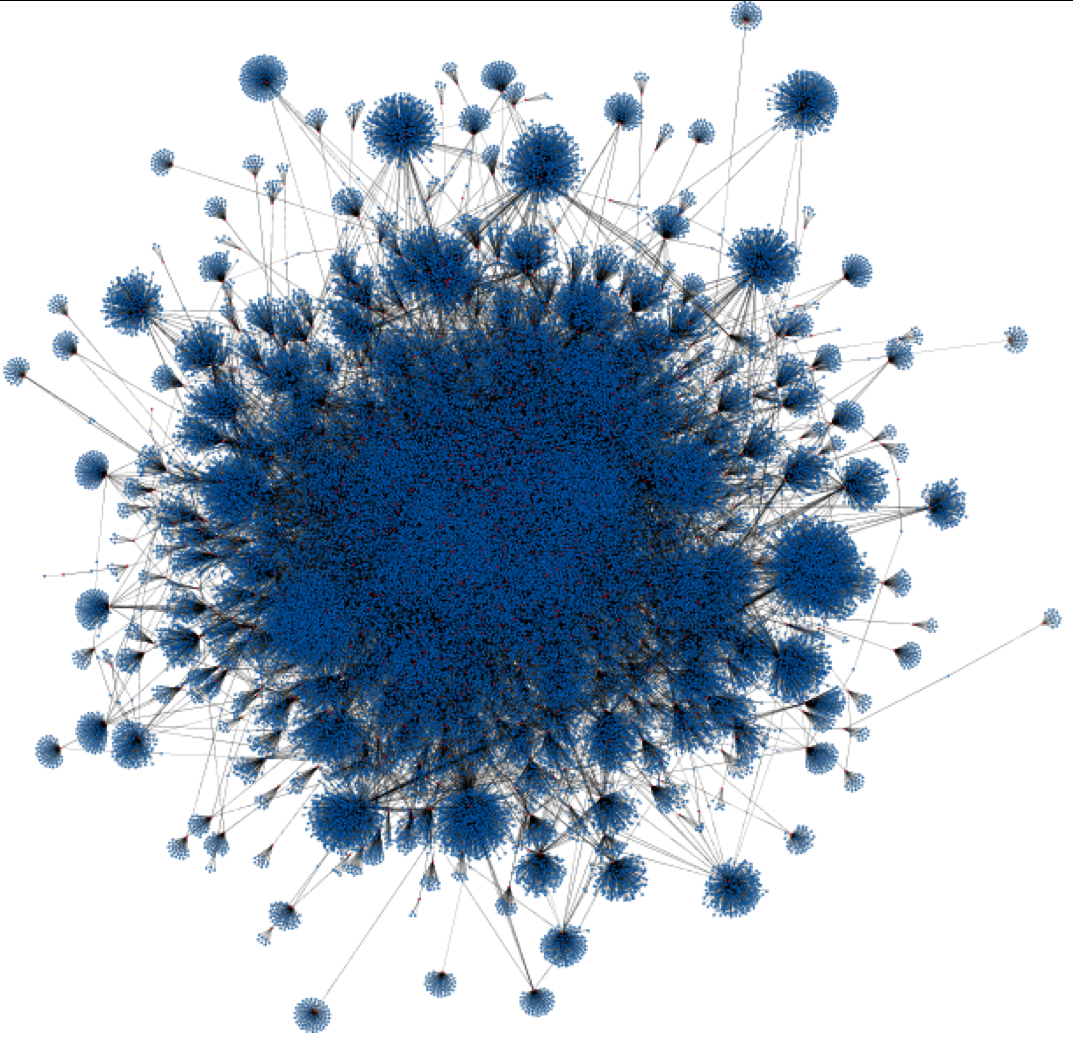}
   }
   \caption{\footnotesize Full Trade-Credit network
   }
   \label{fig:noSnowballsNetwork}
\end{figure}

   The network is directed --- arrows point in the direction of the payments. The
   full network looks like dandelions (many buyers per supplier). An image of the network is displayed in figure \ref{fig:noSnowballsNetwork}.
%   No short paths are evident \hl{SHOW PLOT}.

\begin{figure}[htb]
   \centering
   \includegraphics[scale=0.3,bb=10 10 512 490]{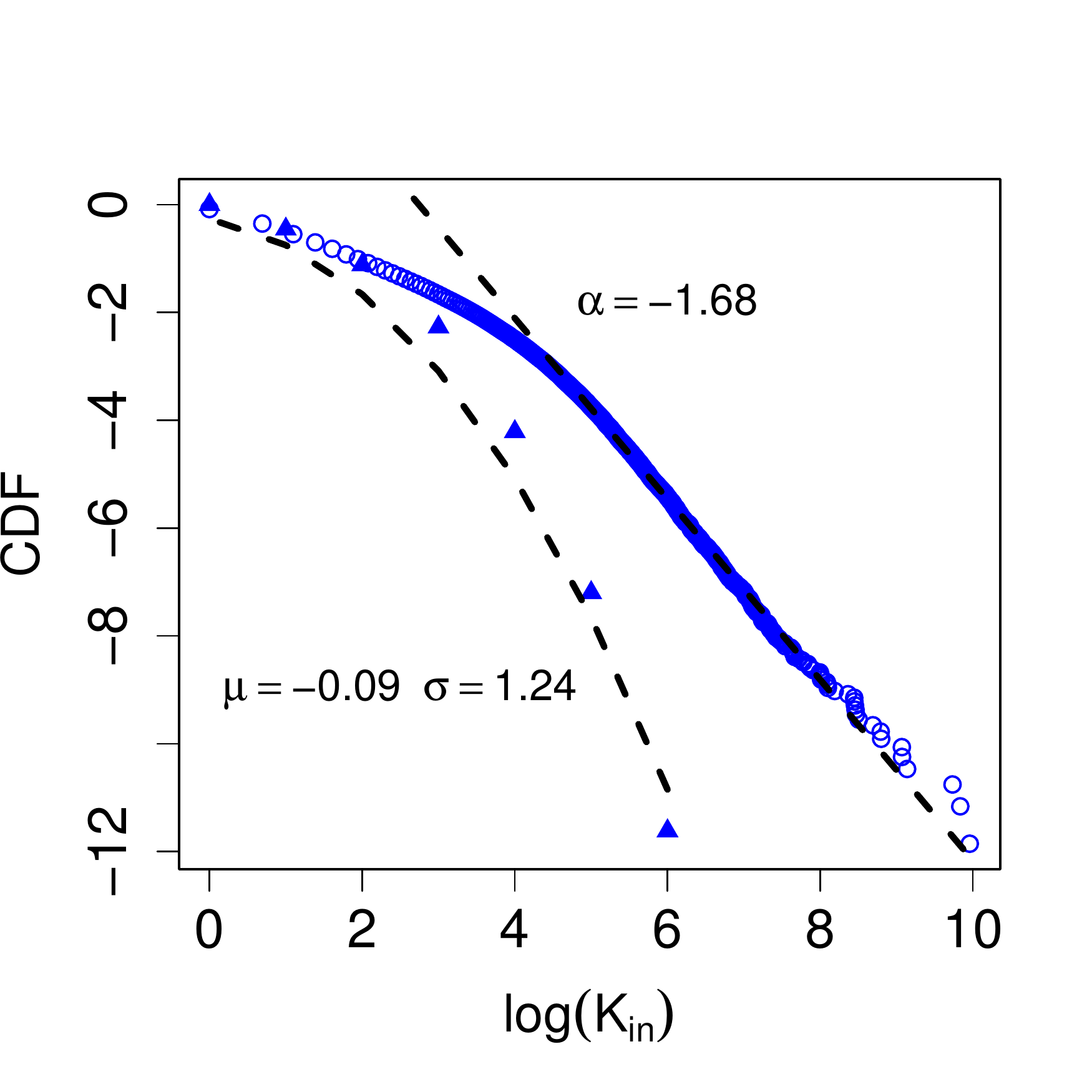}
   \caption{\small Cumulative distributions of firms' in-degree
   (circles $N=129,584$) and out-degree (triangles $N=345,403$).
   %Suppliers that have buyers with full information (TC+BS network, $n=140,580$).
   The reference lines are a power law and log-normal respectively.}
   \label{fig:Kin}
\end{figure}

\paragraph{Degree distributions}
   The in-degree distribution forms a power law over 6 orders of
   magnitude (figure \ref{fig:Kin}),
   the out-degree distribution is a curved line (can roughly be fitted with log
   normal), as can be expected in many kinds of networks where the
   generating mechanisms work on the in-degree.

 \paragraph{Credit Rating}
  A computerized system
  automatically assigns a RATING score to each firm with a
  balance sheet.
   The score is in the range $1\dots 9$ where low credit-rating is indicated by a high value. We can
  further group this index into classes: `A' (RATING $= 1\dots3$) for high
  investment grade firms, `B'  (RATING $= 4\dots6$) for speculative
  and `C'  (RATING $=7\dots9$) that represents firms in risk of default. Firms
  that score into the last class are regarded by practitioners as having
  little or no access to bank credit.
  A firm with a score of 9 will rarely qualify 
  for borrowing. However, since these firms appear  in our TC data as borrowers from the bank,
  we assume that they did receive loans;  a low RATING could
   be caused by the type of industry in which the firm operates. 

  The calculation of the RATING score is proprietary but shows correlation with
  Altman's Z-score \cite{Altman:1968ys}. A comprehensive explanation
  of the RATING score appears in Bottazzi et. al. \cite{Bottazzi:2011fk}.

  It is, however, important to note that the RATING score of a firm
  is visible to all the banks with whom that firm does business.
  This is part of a transparent national credit system that was
  erected in Italy. A common credit registry is also available in
  other countries.
 
\begin{figure}[htb]
   \centering
   \subfigure[frequency]{
   \includegraphics[clip, scale=0.32,bb=5 70 457 415]{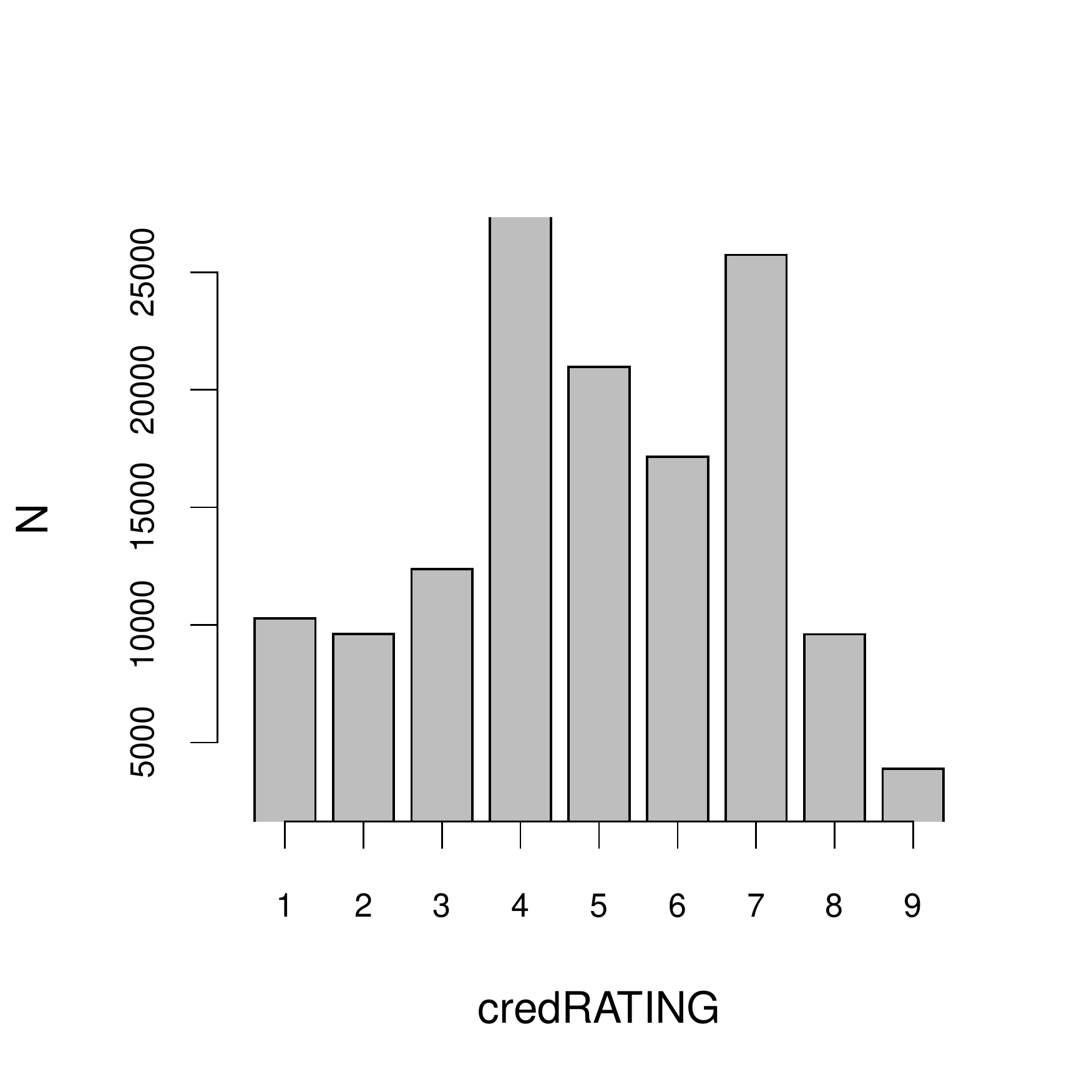}
   \label{subfig:histNumRATING}
   }
   \subfigure[size]{
   \includegraphics[clip, scale=0.32, bb=5 70 457 415]{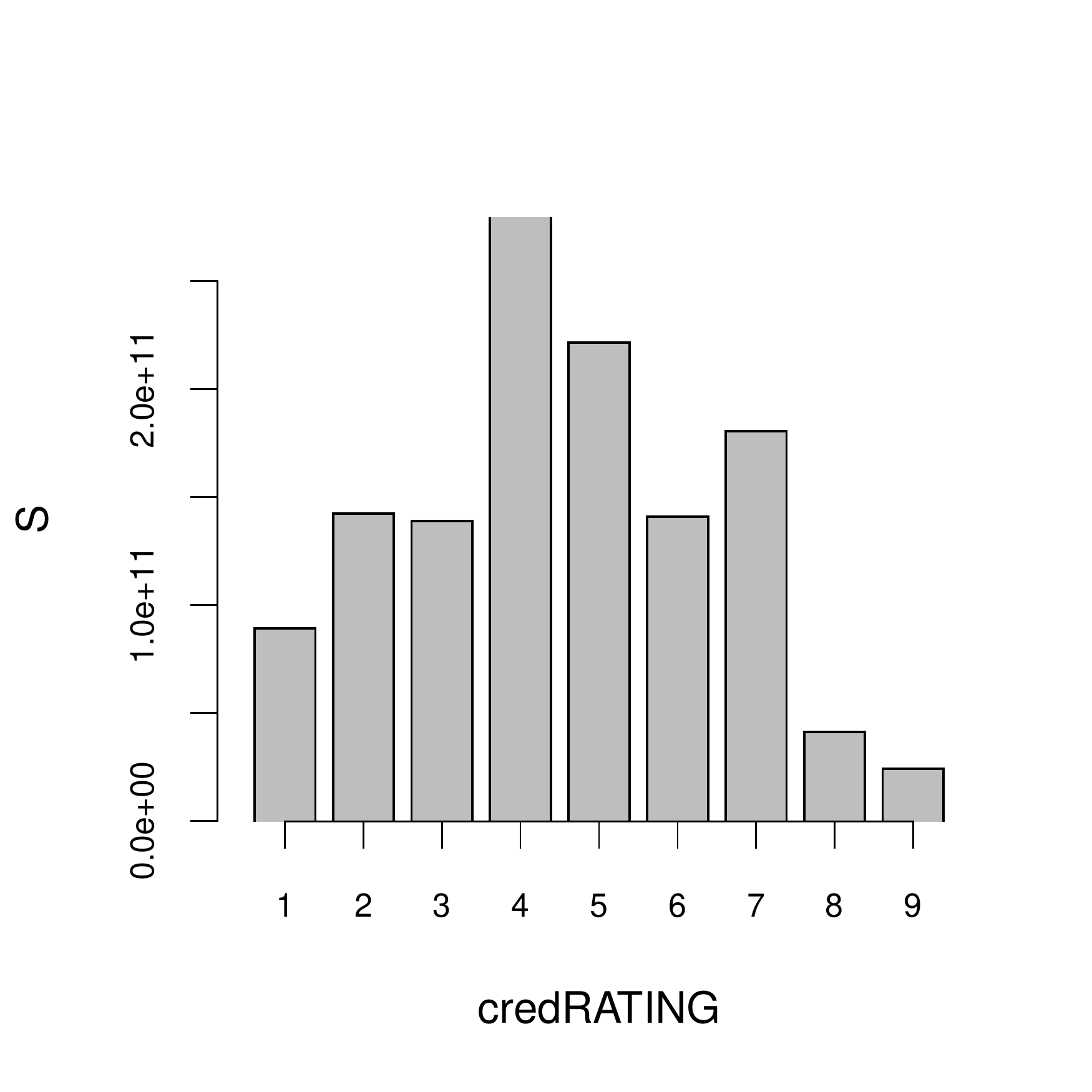}
   \label{subfig:histSizeRATING}
   }
   \caption{\small Histograms of suppliers in \MAll: firm counts
   (N) and total net-sales (S) per RATING score.
   RATING=1, 2, 7 and 8 exhibit the largest deviation of sales per number of firms}
   \label{fig:histRATING}
\end{figure}

  Naturally, there are more buyers than suppliers. However, the
  distribution of RATING scores is identical once normalized to the
  total number in each group. Figure \ref{fig:histRATING} displays
  the RATING histograms of suppliers.

  \subsection{Definitions}
  \label{sec:definitions}
  This section covers the declarations of several parameters we use in the analysis. These include
  the \textit{population} of firms,
  the \textit{transaction pair},
  \textit{annual collateral},
  \textit{information exposure},
  and \textit{credit rating}.

  \paragraph{The information exposure}
  Let us begin by locating the quantities that affect the network visible to the bank.

  We introduce here the quantity we term \textit{information exposure} \ab\!. This is the ratio between the  total amount in the invoices that supplier $s$ registered at the bank during the year, over his annual net-sales, as reported in his Profit \& Loss statement. To describe this quantity we use the following parameters.

  We define the ordered pair $(c,s)$ of customer(i.e. the buyer)-supplier
  directional ties; each firm in the pair is a member of
  the population set `\P\!\!':%The set of all customer-seller pairs, $CS$
 % is a subset of the Cartesian product of the population with itself.

  \begin{align}
    (c,s) \in CS \subseteq \P\times\P.
  \end{align}

  The amount that was presented as collateral we call $R_{cs}$. 
  This is the annual aggregate of the invoices presented by $s$ 
  on account of the contracts written by him to his customer $c$. The
  face value of an invoice serves as collateral for a short term
  loan in the `credit line'. The annual total of all the invoices that $s$ presented
  to the bank is:

  \begin{align}
    R_s &= \kern-1em\sum_{\{c : (c,s) \in CS \}}\kern-1em R_{cs}.
  \end{align}

  This amount should be a good proxy for the amount of the total of short term loans
  that $s$ received  to finance production.
 %For  firms running on low profit margins, $R_s$ will
  %be close to the net-sales of $s$.
  
   We can now define the \textbf{information exposure}, symbolized by 
  \ab. This is the proportion of net-sales, $S_s$, of the supplier $s$,
  that he presented as   collateral $R_s$:

  \begin{align}
    a = R_s / S_s
    \label{eq:a}
  \end{align}

  The net-sales is a parameter in the profit and loss statement of
  a firm and is an annual aggregate (flow) as is the enumerator
  $R_s$.  \a is expected to be relatively close to 1
  for  firms running on low profit margins.
  
  The value of \a is greater or equal to zero and can exceed unity.
  There are three possible situations:

  \begin{itemize}
  \item
  \a could be greater than one. A naive\footnote{%
  We assume that our data does not contain traces of illegal
  activity. Otherwise we would have to remove from our dataset firms that have $a>1$.
  } view for why $a > 1$ is that there
  is misalignment between the time-frames in the data; the closing of the audit
  (P\&L statement) and the expiration of all trade-credit contracts
  that were signed in the same year.

  \item
    If supplier $s$ has $a = 0$, the numerator in
    (\ref{eq:a}) vanished. The interpretation is that $s$ is not a
    direct client of the bank. Rather his customer, $c$, is. The customer
    $c$ entered $s$ into the system by executing an outgoing payment
    transaction.

  \item
  When $a \ll 1$, a discrepancy exists between the total collateral
  and the net-sales. This could hint that the production of $s$
  requires loans that are smaller than the sales. It is an indication
  of a healthy use of the credit line.
  \end{itemize}

  In further analysis we will filter out firms having $a>2$.

  Based on the introduction, the RATING score is expected to have
  a non-trivial relationship with information exposure. We will therefore define
  quantities that relate the two, such as the average information
  exposure for each RATING score.
  This is estimated over supplier firms $s$ that have rating score
  $r$. The set of suppliers with rating $r$ is
  \begin{align}
    \mathfrak{S}_r := \{s:\mathfrak{R}_s=r\},
    \label{eq:Sr}
  \end{align}
  where $\mathfrak{R}_s$ is the RATING score of firm $s$. 
  The average information exposure over all firms having the same RATING is
  \begin{align}
    \bar{a}(r) = \frac{1}{|\mathfrak{S}_r|}\sum_{s \in \mathfrak{S}_r} a_s,
%     \bar{a}(r) &= \sum_{i=1}^{20} w_i(r) \cdot x_i(r)
    \label{eq:aveaR}
  \end{align}
  where  $a_s$ is the information exposure parameter of supplier firm $s$.

\section{Results}

  \subsection{Stylized facts}
  \label{sec:Stylized facts}
    The in-degree distribution is a power law, indicating
    an association mechanism similar to popularity. Although
    the supplier is required to recruit the buyers, in the relevant literature
    on diffusion we witness an inevitable
    coupling of mass media (external factors) and word of mouth
    (local factors). A good historical review of market models can
    be found in Goldenberg et. al. \cite{Goldenberg:2000uq}.
%    The outgoing links (buyers) on double logarithmic scales have a visible curvature. We tried to fit it with a log-normal model. Buyers are less favored by the bank, and therefore we can expect large discrepancy in the buyers pool.  Since this analysis should rely on more information we leave this measurement out of the scope of the current document (\hl{SHOULD I CUT?}).

    Looking at other common network measures on the directed network,
    we find that connected component sizes are small and detached
    from one another with no or only links in reverse directions
    connecting them. The size of the giant component is 101,186
    nodes with a diameter of 20. Of the remaining nodes 239,780 are
    situated in small clusters of 1 link each.
    In many cases the central nodes are buyers
    with small contract sizes, e.g. phone companies or couriers,
    each of which is   financially irrelevant to the system.
    Filtering out the irrelevant firms causes the network to break
    completely.

    When counting the directional chains, the length
    of the longest financially relevant chain (A pays B, B pays C,
    C pays D, .... ) is four. However, these chains are rare and
    across the network there are only eight such chains. The rest
    of the chains are smaller. More common are chains with length
    of up to two.
%  The average path length of the trade-network is 5.8, the average supply chain length is 2.

%    Sreenivasan et. al. \cite{Sreenivasan:2004fk} show that the
%    length of the optimal path is expected to be relative to log
%    of the network size $\log(340,000) = 12.7$. Contrast to their
%    priors, here we disregard the weights on the links. Still, in
%    effect the average path length of our directed network is much
%    larger, rendering any flow measurements impossible. I.e. the
%    links over which information can pass efficiently, are not
%    visible to us.

%   The majority of the visible paths will therefore not realize.
%   And among the links that we do observe, it is harder to believe long paths will realize.

%    {\color{blue}Why does an average path length make flow measurement impossible?} \textcolor{red}{because the majority of the paths will not realize in real life \cite{Sreenivasan:2004fk}}.

    The fact that we cannot trace the supply chain was another
    hint that data is severely missing.

%   {\color{blue}If the average
%   path length is 12, why can't the production chains be traced?
%   Also I though the paths lengths were small compared to industry
%   average of 4. Is there a typo here? Maybe  the average path
%   length is 1.2 rather than 12}
    However, there is an excellent financial explanation for this:
    Manufacturers are more interested in making the product, and
    less interested in storing and transporting it. For this task there
    is a dedicated industrial class called `wholesale'. A \textit{Wholesaler}
    will retrieve the goods from the manufacturing plant, and make sure
    it arrives at its destination, maybe a retailer or maybe another
    manufacturer down the supply chain.

    The accounting procedure of wholesale is to sell the product
    on behalf of the manufacturer. It does not have a production
    cycle and acts as a goods buffer.  The wholesaler delivers the
    goods on the condition that he gets paid for them immediately
    when they're sold. Consequently, it serves as a money buffer,
    and there is no contractual date by which the invoice must be
    paid. So, the bank will not accept it as collateral.
%   and transfer the funds to the manufacturer when
%   they are received, deducting compensation for their service. So the invoices
%   are not booked as accounts receivable on the wholesaler's
%   side.

  We make two assertions here: that link information is missing
  from the inter-firm trade network and that wholesale firms break
  the connection downstream along the supply chain.

%    {\color{blue} How does the break occur? The supplier invoices the manufacturer for what he has sold in total over say a month, and the wholesaler invoices the buyer. There is continuity in invoices. Do you imply that the wholesaler does not present the invoices he writes to his buyers to his bank for discounting, which breaks the chain of presented invoices? If so, this needs to be added at this point.}

  \subsection{The characteristics of a supplier's neighborhood of buyers}
  \label{sec:neighborhood character}
\begin{figure}[htb]
   \centering
%   \subfigure[full network]{
   \subfigure[mass]{
   \includegraphics[scale=0.3,bb=10 10 512 490]{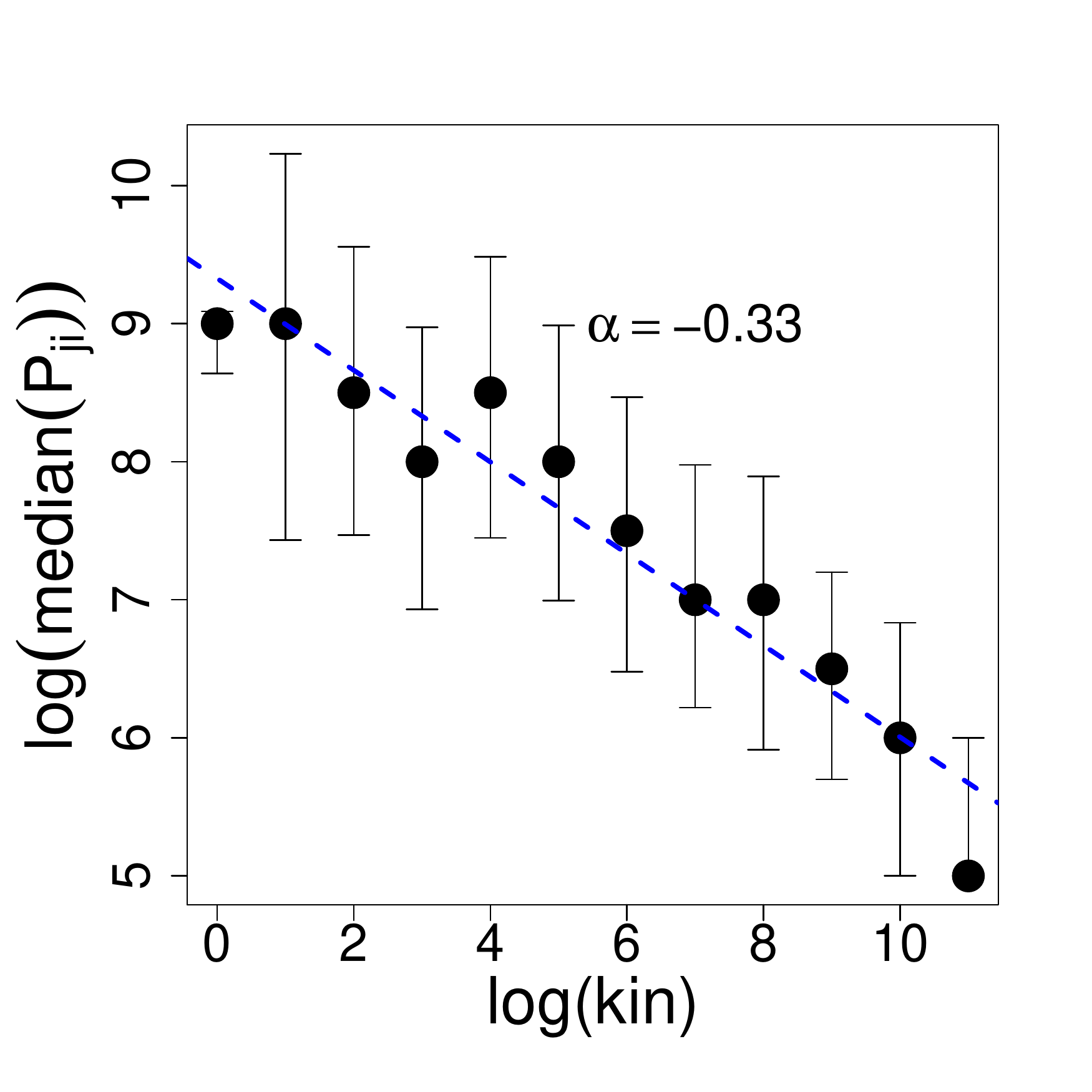}
   \label{subfig:kvsPji}
   }\hfil
%   \subfigure[$\MAll$]{
   \subfigure[number]{
   \includegraphics[page=3,scale=0.3,bb=10 10 512 490]{run\mydot UO4m3/knn_CDF\mydot in\mydot 0_00\mydot UO4m3.pdf}
   \label{subfig:knn}
   }
   \caption{\small 
   Median contract size \ref{subfig:kvsPji} and average buyer
   (neighbor) degree \ref{subfig:knn} plotted
   against supplier's degree ($K$ or $K_{in}$).
   Taken from the suppliers in the trade-credit full network (%\ref{subfig:kvsPji}
   n=273,726).
%   and the restricted to the seller firms identified with balance sheets
%   (\ref{subfig:NjvsPji} n=129,584).
   The network is dissortative both in connectivity and mass.
%   median contract size and neighbors linkage drops with connectivity.
   The fitting line on panel \ref{subfig:knn} has exponent $\alpha=-1.246$ }
   \label{fig:kvsPji}
\end{figure}

    Generally, large firms deal with large contracts, small firms deal with small ones,
    although there are exceptions, e.g. the phone and courier companies.
    From figure \ref{fig:kvsPji} we learn that a large number of
    clients means both that the median contract size is small and
    with high probability the neighbor's degree is small. This
    is an indication that
    clustering around a supplier is negatively assorted. In other words,
    highly connected firms tend to be positioned far from one another
    in the network and thus render the network more vulnerable to
    systemic shock (removal of the highly-connected suppliers creates
    an impact across the whole network because their neighborhoods
    are not densely interconnected). However, the network is less
    likely to percolate in the sense that distress does
    not spontaneously amplify itself \cite{Newman:2002fk}.

\begin{figure}[h]
   \centering
   \subfigure[difference per RATING]{
   \includegraphics[clip,scale=0.3,bb=70 410 495 750]{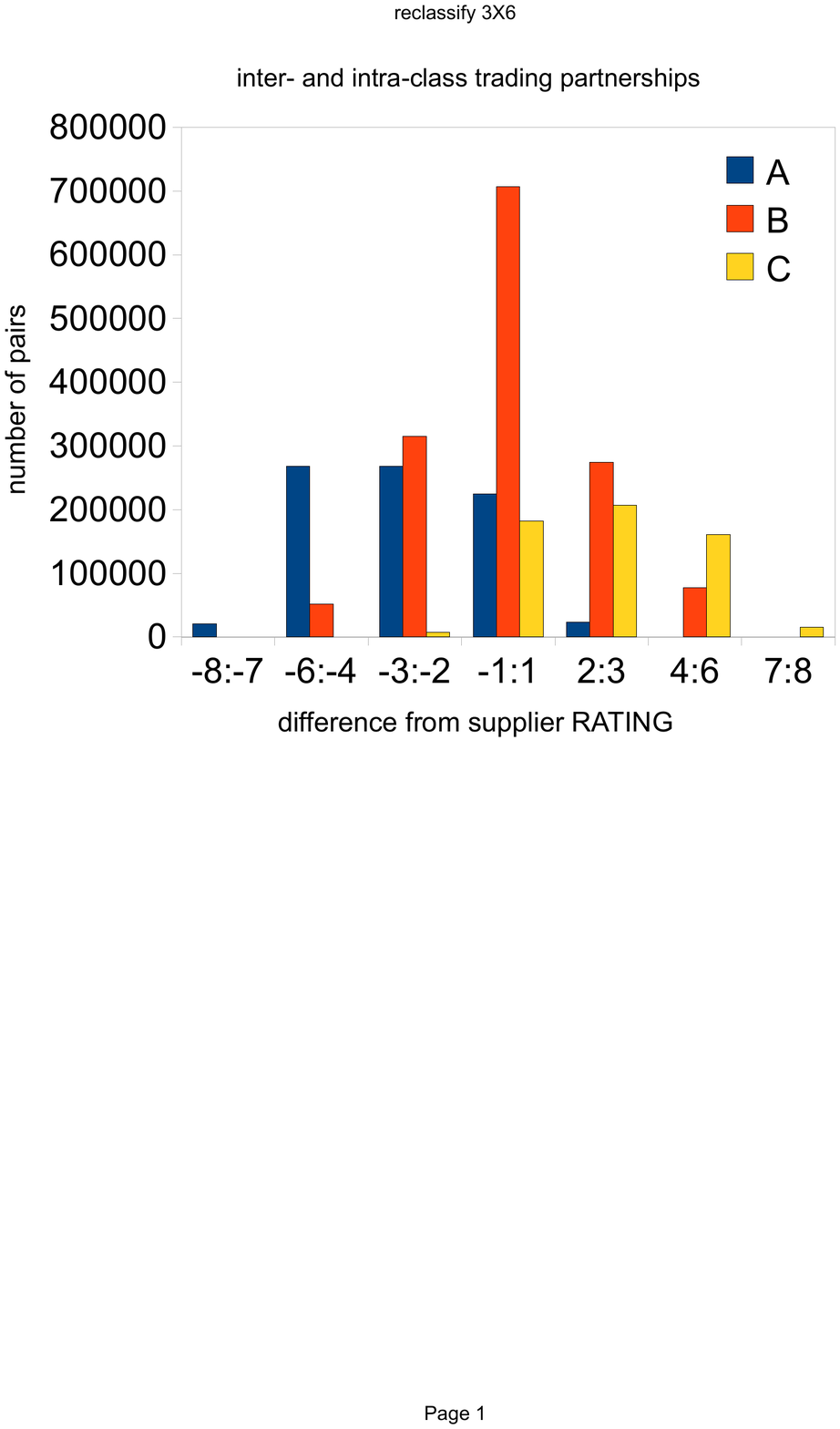}
   \label{subfig:homophily RATINGd}
   }\hfil
   \subfigure[mosaic]{
   \includegraphics[scale=0.28,bb=40 65 450 420]{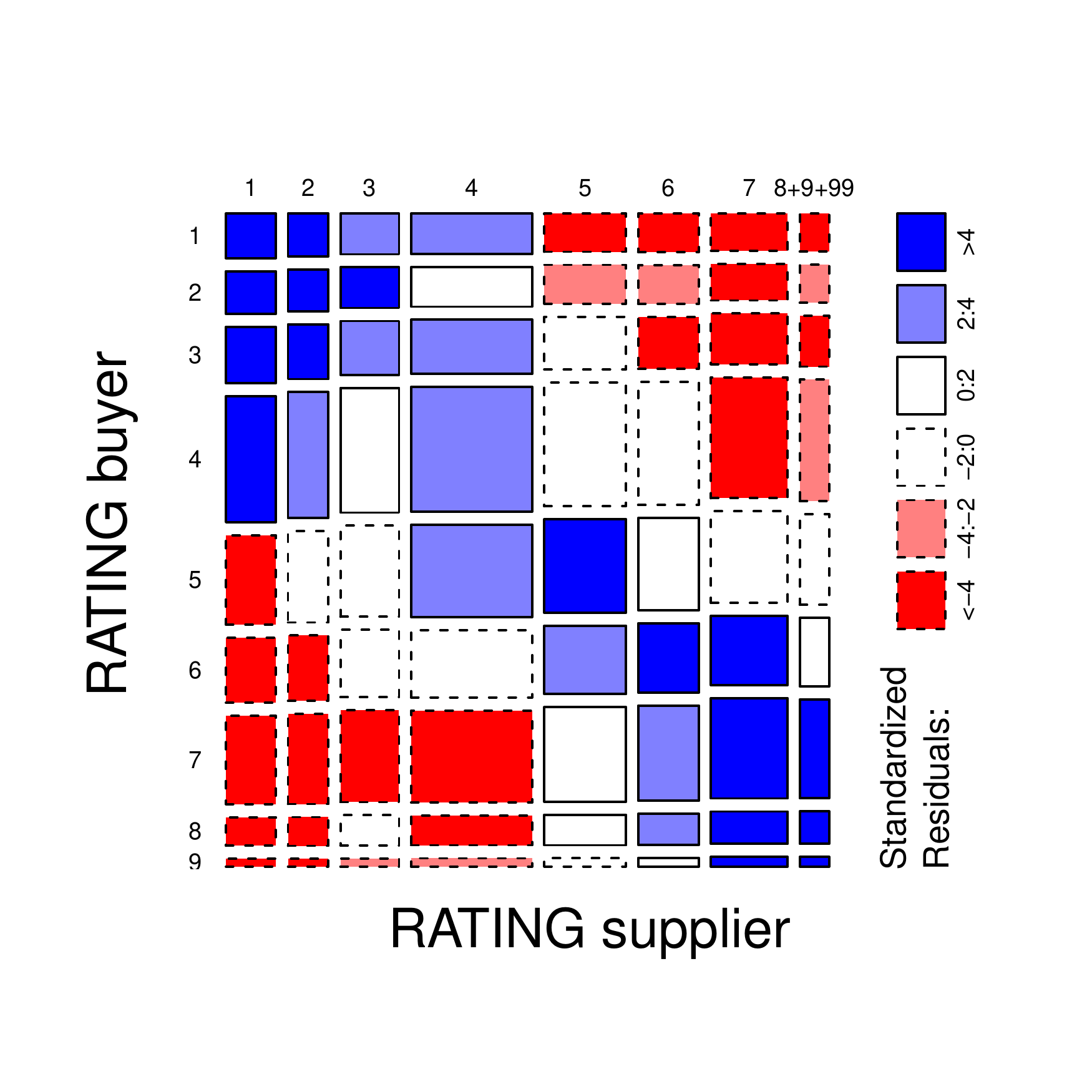}
   \label{subfig:homophily RATING mosaic}
   }
   \caption{\small Affinity between suppliers and buyers. The X-axis
   in panel \ref{subfig:homophily RATINGd}
   gives the difference supplier RATING minus buyer RATING. The Y-axis
   counts the number of pairs with difference categorized by X. The
   suppliers are classified into `A' `B' or `C' class, and the
   differences are grouped every 3. There are 2,802,976 pairs in total.
   Panel \ref{subfig:homophily RATING mosaic} shows the compressed
   tiles mosaic of the cross tabulation RATING supplier $\times$ RATING
   buyer. The plotting scheme is described by Friendly \cite{Friendly:1994uq}
   }
   \label{fig:homophily plot}
\end{figure}

 We created a cross tabulation of RATING scores for all
 suppliers in data set $\MAll$ and their buyers that have RATING
 information (TC+BS), which is summarized visually in figure~\ref{subfig:homophily RATING mosaic}.
 The RATING of the supplier is in the columns and the RATING of the buyer
 in the rows. The table holds the counts of all pairs of RATING
 scores possible in the data. Essentially this is a description of RATING
 on the two ends of each link between trading firms.

 The number of pairs where the supplier has RATING score greater than
 8 were aggregated, leaving a total of 8 groups of supplier
 RATING scores. The RATING scores of buyers remain a total of 9 RATING
 groups.

 From this table we estimated a $\chi^2$ test of independence of
 the categories. This test produced a statistic $\chi^2 =
 2803$ with 56 degrees of freedom and a p-value  identical to zero.
 The conclusion is that we can reject complete independence
 between RATING of supplier and buyer and suggest a tendency of suppliers
 to affiliate with buyers having a similar RATING. Panel \ref{subfig:homophily RATING mosaic}
   shows the tile mosaic of the paired RATING classes
   with color coding that reveals this tendency;
   the area of each tile in the mosaic is proportional to the
   number of pairs where supplier has RATING=X and buyer has RATING=Y.
   A blue color marks significantly higher than expected
   occurrence, and a red color paints a significantly lower than
   expected pair count.
 
   It is important to note at this stage that
   a $\chi^2$ test of independence is categorial and does not take
   into consideration any ordering of columns or rows. However, the
   table used as input maintains the original ordering in both
   dimensions; the RATING classes.
   Thus, the pattern that appears as blue along the diagonal
   does indicate higher-than-expected encounter of similarities in
   the two nodes sharing a trade-link. And any other ordering of the
   columns or rows would result in a less compelling pattern.

\begin{figure}[htb]
   \centering
   \includegraphics[clip, scale=0.32, bb=40 370 510 730]{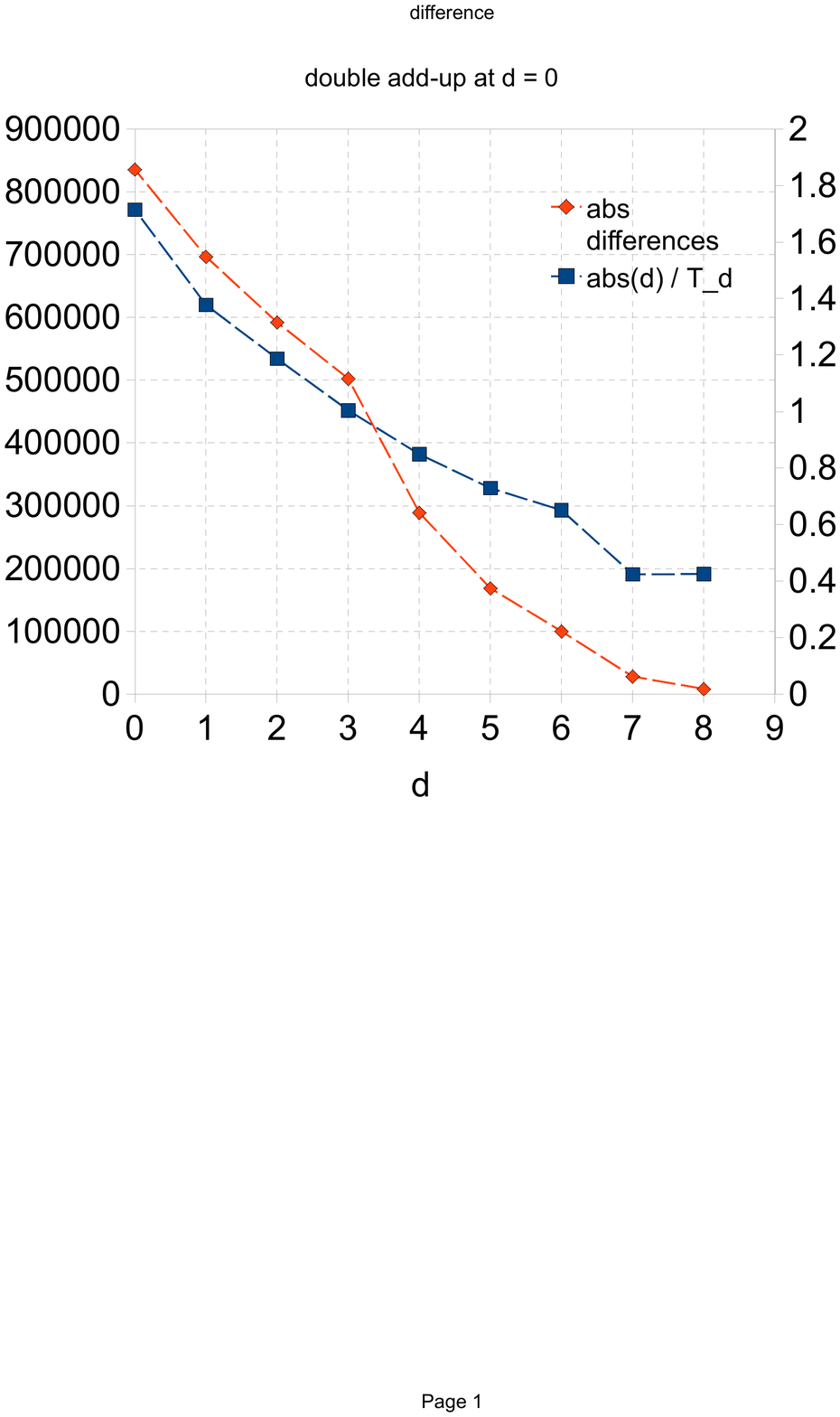}
   \caption{\small Histogram of
   absolute RATING difference $d$: The left Y-axis gives the number of pairs (supplier-buyer)
   that have an absolute  difference in RATING of $d$, corresponding to the red
   curve. The right Y-axis is for the corrected counts (in blue). The plot
   shows both (1) that the number of links (between suppliers and
   buyers) decreases with difference $d$. (2) that the number of links
   (between suppliers and buyers) adjusted (for max possible) also
   decreases with difference.
   }
   \label{fig:marginals}
\end{figure}

    The plot in figure \ref{subfig:homophily RATINGd} shows
    that buyers have similar RATING scores as their suppliers. In
    figure \ref{fig:marginals} we see another indication that the
    probability of pairing with different RATING scores is decaying
    with the difference in RATING score. The decay is linear up to
    three RATING notches away, then the drop becomes sharper both in
    slope and in shape. This is a hint that members within the same
    major rating class (1..3, 4..6 or 7..9) differentiate by the
    distance of RATING scores.

\begin{figure}[htb]
   \centering
   \includegraphics[scale=0.3,bb=40 60 480 490]{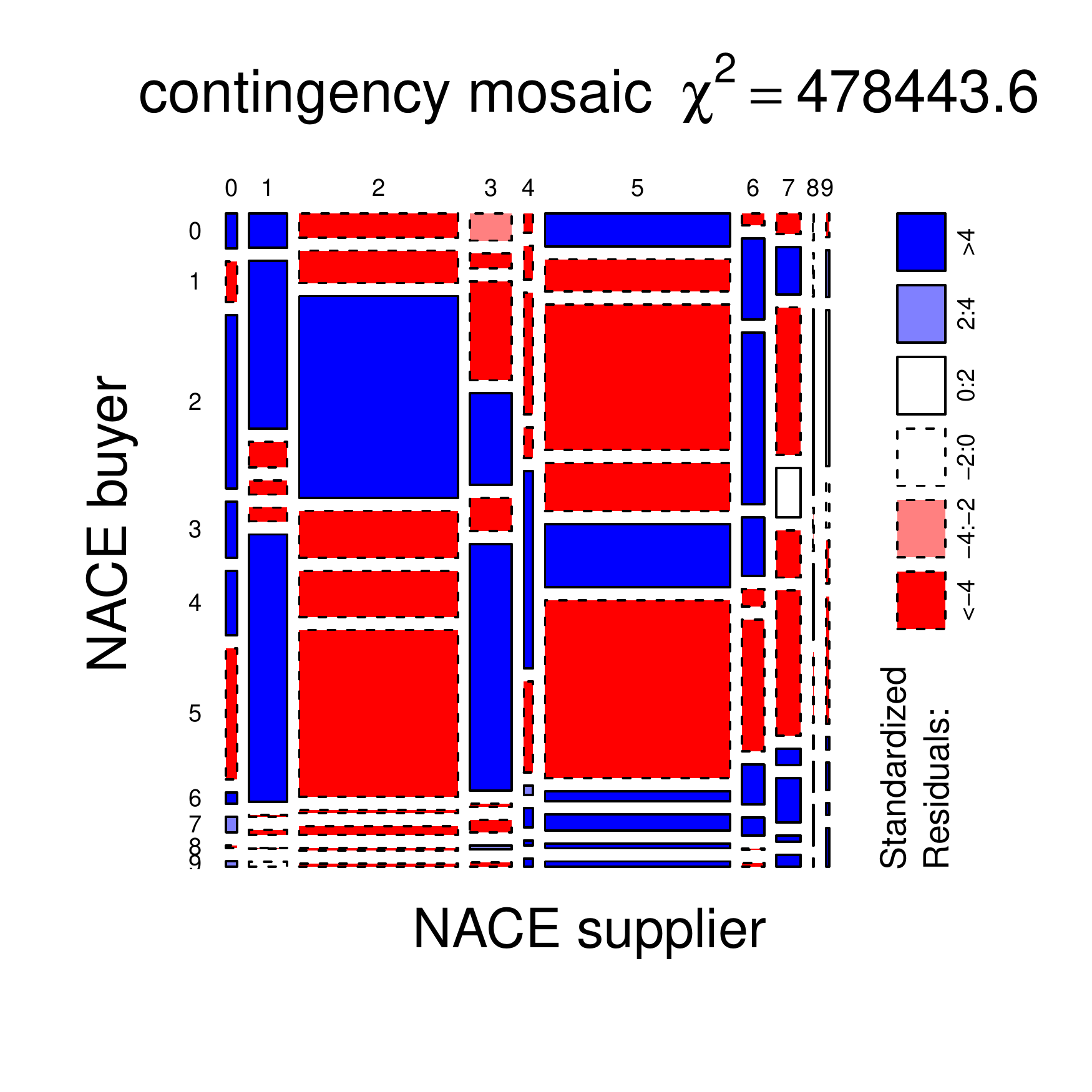}
   \caption{\small Input-Output test of independence between buyer
   and supplier. 1-digit NACE industrial classifications are roughly:
   $1\dots3$ manufacturing, 4 is energy, 5 is wholesale and retail,
   6 is transport, and 7 is real estate.}
   \label{fig:IO}
\end{figure}

   One reservation could be made on the result above: it is remarkable
   that the RATING of the supplier is so similar to the RATING
   score of the buyer. Looking at the sectoral affiliation of the
   buyers and the suppliers (figure \ref{fig:IO}) it seems that
   suppliers and buyers are, in the main, trading inside the
   same industries especially when within the manufacturing sectors
   (NACE categories 1 .. 3).

   The visual mosaic is symmetric, i.e. it can be
   transposed while maintaining an almost identical pattern of red and
   blue tiles. One exception is NACE major category 5, wholesale, that disrupts this symmetry.
   Firms in industries 1 and 3 sell to firms in 5. According to our
   data set firms in 5 do
   not sell to those in 1 nor 3, but rather to those in 4 (energy).
   In the real world, wholesale trade does connect between manufacturers, but in our case, it only provides a transient path out
   of the manufacturing industries, and splits the supply chain.
%   {\color{blue} ADD IF CORRECT otherwise supply another explanation: Category 5 is wholesale firms.}
   As explained above (\ref{sec:Stylized facts}), this is due to the accounting procedure of
   wholesale and retail firms. Now, if a node is removed off the tree
   subgraph identified as the supply chain, it inevitably
   lengthens the travel across the network.  This result
   supports the long path length result obtained in \ref{sec:Stylized facts}.
%   Wholesale is in essence storage and transport of goods. These
%   firms store and sell the goods in behalf of the manufacturer,
%   and thus don't buy on credit.

   Since sectoral affiliation is known
   as one of the macro-determiners of RATING scores, we addressed this
   conflation by creating a list of firms that trade outside their industry, and then performing the test again.
   Approximately 15\% of the buyer-supplier links connect firms from
   the same 2-digit industry classification. Figure \ref{fig:NACEiNACEj}
   shows the test of independence of RATING scores without the
   intra-industrial trade links. The table attached above the figure
   shows the sizes of the sets. The test statistic is $\chi^2 = 1456.1$
   with 49 degrees of freedom and $p$-value of zero, again rejecting
   the hypothesis of complete independence and suggesting a trending
   behavior; the firms that a firm trades with have a RATING score
   that is similar to its own RATING score.
   % It is now more obvious also that RATING category 4 (which is large by counts) is exceptional in the sense that they affiliate with buyers of higher credit-rating.

    \pgfplotstableread[row sep=newline,col sep=comma,header=true]{
classification,ieqj,inej
1-digit,873485,2001345
2-digit,449012,2425818
            }\NACEijtable
            %total 2874830

    \pgfplotstableread[row sep=newline,col sep=comma,header=true]{
classification,ieqj,inej
1-digit,873485,2001345
2-digit,449012,2425818
            }\NACEijtable
            %total 2874830

\pgfplotsset{anchor=center,/tikz/baseline}
\tikzset{imagenode/.style={anchor=center}}

\begin{figure}[htb]
     \centering
  \subfigure{
\resizebox{0.65\linewidth}{!}{%
 \pgfplotstabletypeset[
  col sep=comma,
  columns={classification,ieqj,inej},
  int detect, 
  columns/classification/.style={string type, column type={|c}},
  columns/ieqj/.style={column name=$\mbox{NACE}_i \eq \mbox{NACE}_j$, column type={|c}},
  columns/inej/.style={column name=$\mbox{NACE}_i \ne \mbox{NACE}_j$, column type={|c|}},
  every head row/.style={before row=\hline,after row=\hline},
  every last row/.style={after row=\hline},
]{\NACEijtable}
}
\label{subfig:NACEijtable}
}
\hfill
  \subfigure{
  \begin{tikzpicture}
    \node[imagenode] (image1) {\includegraphics[scale=0.35,bb=40 65 450 420]{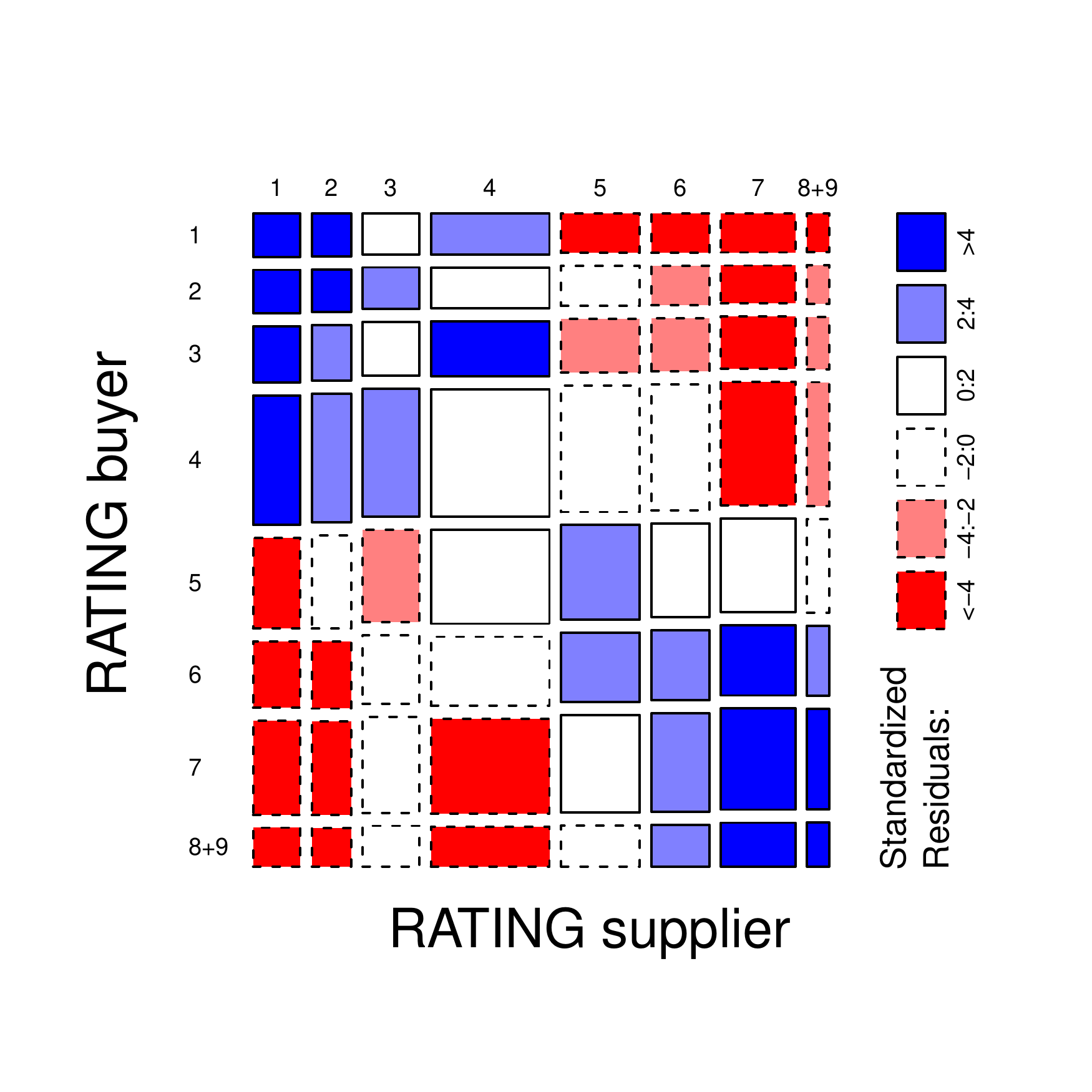}} ;
  \end{tikzpicture}
   \label{subfig:homophily RATING mosaic NACEij}
  }
  \caption{\small The table on the top %\ref{subfig:NACEijtable}
  lists the number of buyer-supplier pairs from the same
  industry, and from other industries. NACE is the industrial
  classification code of the firm. On the bottom
 % Panel \ref{subfig:homophily RATING mosaic NACEij} gives
  is the compressed tiles mosaic for the $\chi^2$ test of independence of RATING
  between inter-industry buyer and supplier. After removing
  the RATING=99 on either end there are 2,325,928 pairs. Compare
  to \ref{subfig:homophily RATING mosaic}
 }
  \label{fig:NACEiNACEj}
\end{figure}

  \subsection{Analysis of the information exposure parameter}
\begin{figure}[htb]
   \centering
   \subfigure[ ]{
      \includegraphics[page=3,scale=0.35,bb=10 30 500 470]{run\mydot 8760b/a_vs_iRATING\mydot 8760b.pdf}
   }\hfil
   \subfigure[ ]{
\begin{tikzpicture}[
        every node/.style={anchor=south west,inner sep=-30pt,scale=0.5},
        x=1mm, y=1mm,
      ]   
     \node (fig1) at (0,0) %
%     {\framebox{\includegraphics[scale=0.75,bb=40 35 450 500]{run\mydot f399a/MDgt0_10X10\mydot f399a.pdf}}};  
     {\includegraphics[scale=0.75]{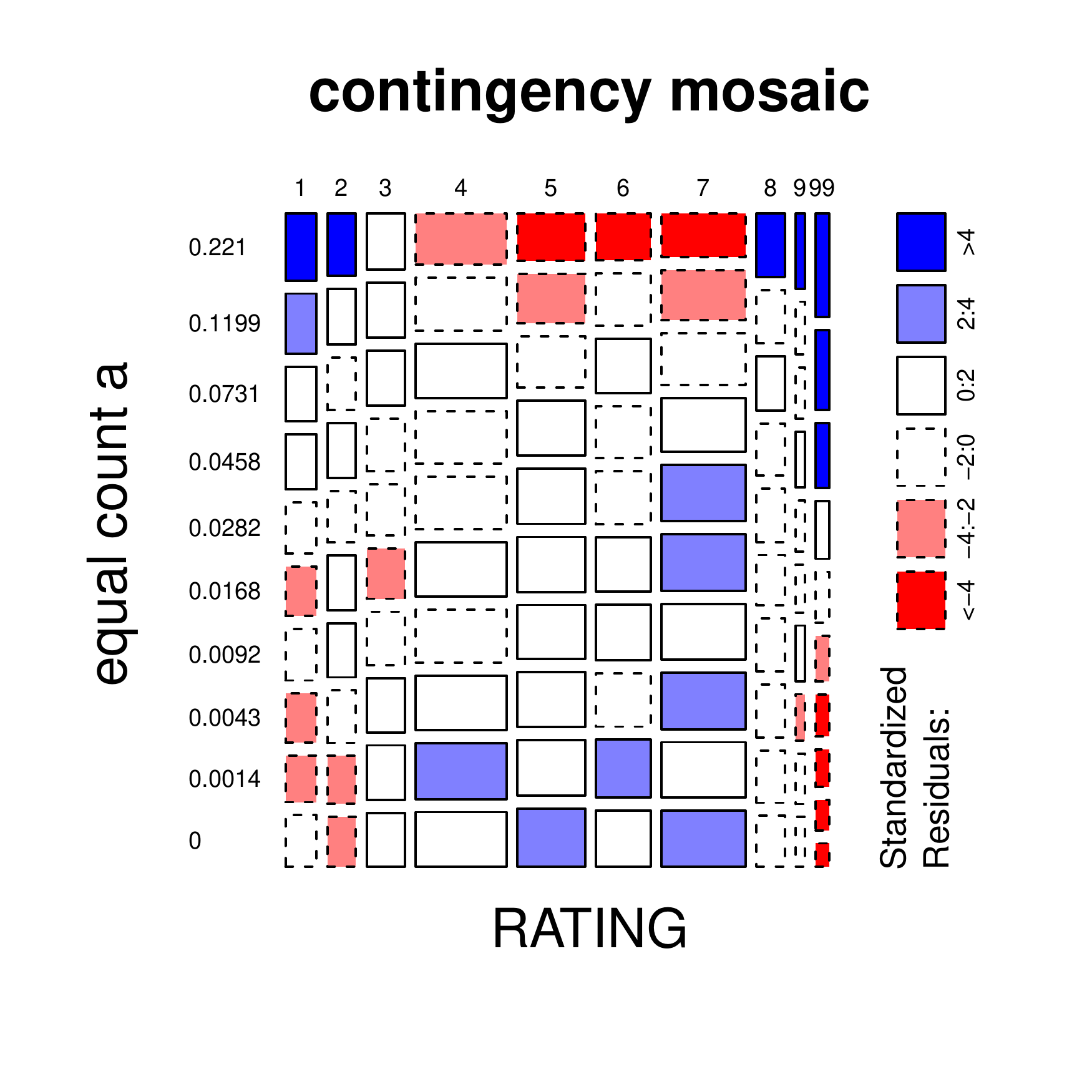}};  
     \node (fig2) at (18,15)
     {\includegraphics[scale=0.75,bb=90 115 450 300,clip,width=170pt,height=220pt]{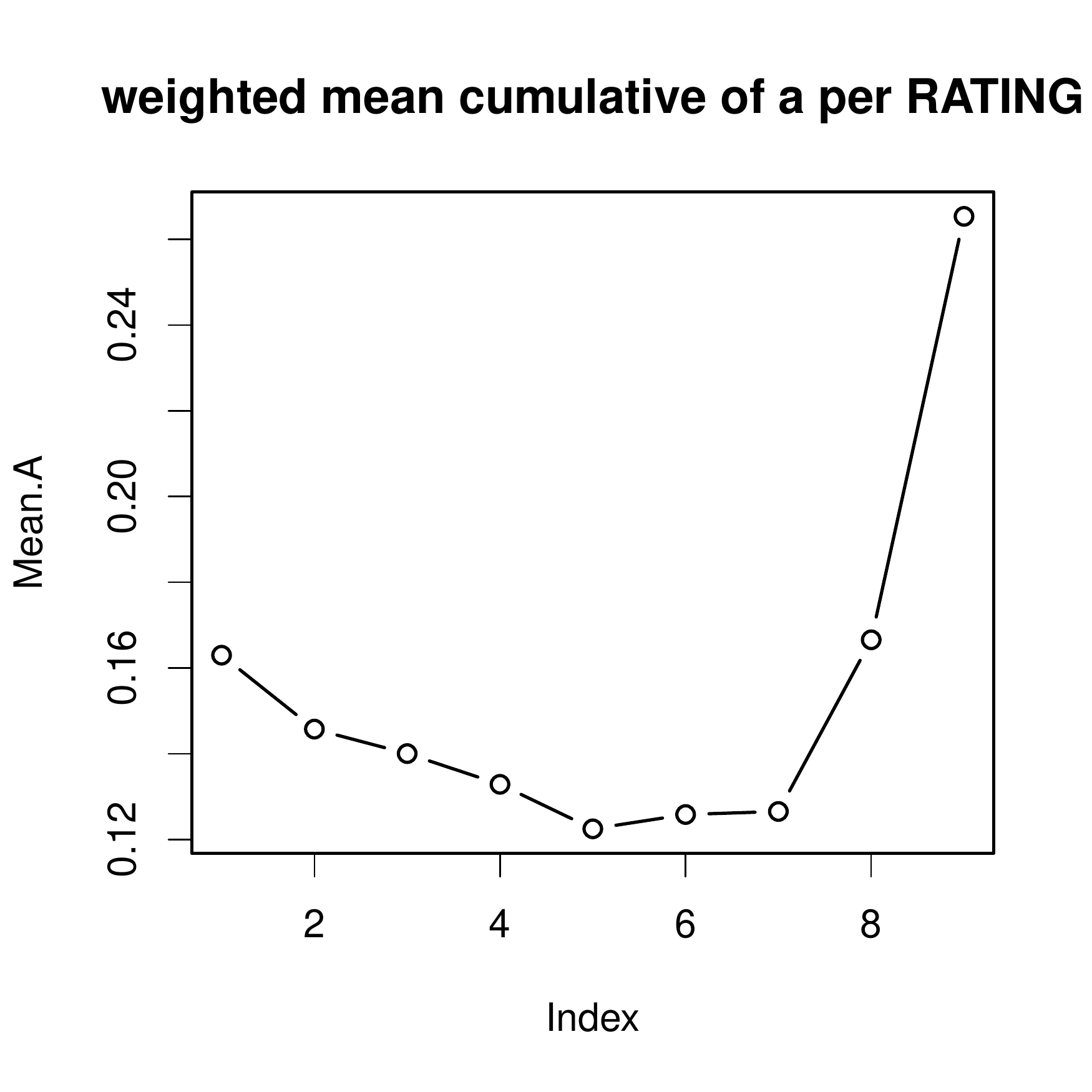}};
\end{tikzpicture}
   \label{subfig:meanAvsRATING mosaic}
   }
   \caption{\small RATING group $r$ vs. average information exposure
   parameter $\bar{a}(r)$ as defined in (\ref{eq:aveaR}). The bottom panel
   is a mosaic plot of the contingency table used for
   the statistical analysis. The grouping procedure is described
   in the text. To illustrate the U-shape there is a line
   plot of panel (a) stretched over panel (b)}
   \label{fig:meanAvsRATING}
\end{figure}

   Figure \ref{fig:meanAvsRATING} shows the relation between RATING and the average information exposure, \a, in two different forms. From this figure we see that the average information
   exposure is at its minimum in the middle of the RATING scale.
   RATING scores of firms in the `speculative' financing group
   $4\dots7$ have the lowest average information exposure.  These
   firms in the middle are believed to optimize
   the amount of information they expose.

   The distributions are skewed, so using average
   \a may not be so informative. For this reason we created a second
   partition of the data set. This is a division into equal-count
   \ab-groups: the set of 129,584 firms was ordered
   by \a and then a division into groups was made every 12,958 or 12,959 records. The
   lowest value of \a in each group
   is placed on the Y-axis of panel \ref{subfig:meanAvsRATING mosaic}.
   The area of each rectangle in the mosaic is proportional to the count
   of sellers that have RATING=X and $a \in [Y,Y+1)$. The color code marks
   either significantly higher (blue)
   or significantly lower than expected (red) frequency of occurrence.

   The effect of RATING on \a is a U-shape visible in both panels.

   In the two extreme RATING scores, 1 and 9, the information exposure is
   the greatest. This sits well with the expectations that `investment' grade
   firms will have dispersed their risk and therefore are
   indifferent to collateral quality. And that firms in risk of default
   will be (or think they are) forced by the bank to surrender all
   possible information.

\begin{figure}[htb]
   \centering
   \includegraphics[scale=0.32, clip, bb=0 50 470 440]{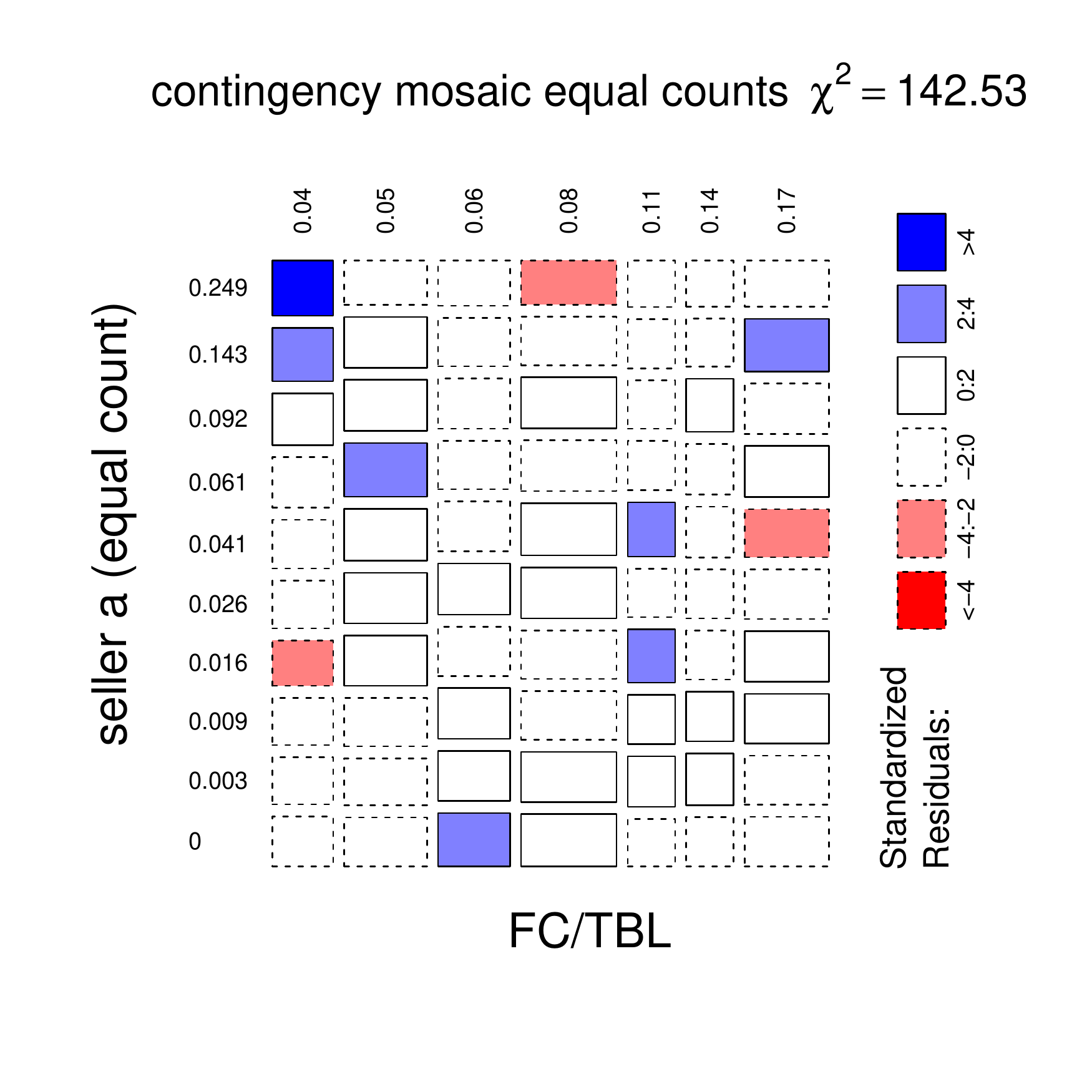}
   \caption{\small Cross tabulation of the Financial Costs (interest
   paid) over Total Bank Loans vs. \a in data set $\MAll$. The test
   suggests three regimes of high correlation: low information
   exposure firms with medium interest, and high information exposure
   with either high interest (low-credit rated firms) or low interest
   (highly credit-rated ones).
 The test of independence gives $\chi^2 = 142.53\ df=54$.}
   \label{fig:FCTBLvsA}
\end{figure}

  Last, we want to show that the information exposure \a is correlated
  with the interest paid by borrowers. Figure \ref{fig:FCTBLvsA}
  is a cross tabulation mosaic of Financial Costs (normalized by
  total bank loans) in the columns, and \a  in the rows. A $\chi^2$
  test of independence was performed and complete independence was
  rejected. We can observe that low information exposure is uniquely
  identified with borrowers that sit in the center of the interest
  scale. There are two more populations of borrowers that have high
  information exposure: firms that paid the lowest interest, and
  these that paid the highest.

\section{Conclusion}
  The trade-network is negatively assorted in respect of degree correlations, but positively assorted when looking at the relationships between attributes of neighboring nodes.
%  The data was probed with a selection bias and
  The \textit{credit-rating} parameter is important to the borrower
  firm and the bank. In the financial sense it represents the future ability of a firm to
  finance its production or purchases, and implicitly it represents
  the ability of a node to associate with other nodes: with high
  probability, firms of similar rating will tend to associate. Medium credit-rated suppliers will selectively expose information about their transactions with neighbors while firms that sit on the ends of the rating scale (high or low) will be less restrictive in exposing information on their neighbors.
%  associate with buyers of their own rating class or higher.

%  To greater extent, medium credit-rated firms control the exposure
%  of information on buyers to the bank.
%  By doing so, they
%  reduce the risk of depreciating the terms on loans - which result
%  in lower interest on the loans.
   In the case of medium credit-rated borrowers the social component
   is related to an accounting behaviour on part of the borrower
   firm that is fondly termed in the literature `window-dressing'
   \cite{Allen:1992fk}.

%  In our data set the \textit{credit-rating} is a result of the interaction between the nodes in the system and the probing procedure of the bank. 
  In the network that we observed, a large span of data is missing.
  The reasons for missing information could be divided into two:
  \begin{itemize}
    \item nodes which are suppliers, are missing due to a selection
    bias driven by business preferences of the bank.
    \item nodes which are buyers and links from suppliers to existing
    buyers, are missing because of the interaction between the bank
    and the supplier (borrower).
  \end{itemize}
  
  We were able to: \textbf{(a)} show that data is missing but not
  at random and suggest that the data collector (bank) has an impact on the process by which firms tend to associate, and especially on the medium
  credit-rated classes; \textbf{(b)} show that the immediate
  neighborhoods of firms from the medium credit-rated classes are
  more likely to be missing complete links and nodes structure;
  \textbf{(c)} suggest a tendency of this network to be vulnerable
  to targeted removal of specific types of nodes. These types may be
  correlated with industries, much like the scenario of a systemic
  shock where the bank issues a regulatory action on the industry
  as a whole.

%  \textcolor{red}{%
%  Minsky's moment theory \cite{Knell:2012vn} describes the business
%  cycle as a periodic shift between stable and unstable states of
%  the economy. Two factors cause the shift: market internal
%  dynamics and the intervention of a regulatory system. Agent based models are able to simulate the internal dynamics of the economy and support it \cite{Lorenz:2009kx}.

  Since the bank is collecting the data and so assumes a point of view
  we suggest that \textbf{(d)} the activation of a large scale
  distress response due to internal dynamics is rarely visible to
  the bank. The synchronized response subsequent
  to an intervention of the regulatory system is nevertheless
  visible, due to the causal nature of it.
%
%  . Thus we suggest that \textbf{(d)} financial crises are
%  rarely caused by a local distress that spreads and bubbles up
%  spontaneously \cite{Lorenz:2009kx}. More likely is a
%  synchronized response in many locations that occurs subsequent
%  to a systemic shock (regulatory action).
%}

  From these findings we are able to conclude for the Italian economy whenever
  there is a financial crisis in the banking system that:
  \begin{enumerate}[(a)]
    \item The firms with a poor RATING score will be severely hit by
    their bank's refusal to extended a credit line on their trade
    bills since they depend heavily on the bank to do so (their
    information exposure measure, `a', is high.
    \item If the banks also refuse trade credit to the firms with
    good RATING score, they too would have to find other sources to
    finance their continued production; how they might do this would
    be pure speculation on our part.
    \item If the banks reject trade credit to middle RATING firms,
    these firms would not be so badly affected as they present a
    smaller proportion of their sales invoices for discounting.
    \item From a national perspective, it may be in the government's
    interest to encourage banks to continue to extend credit to the firms
    with poor RATING as these are the more likely to go under if
    their trade credit applications are refused.
  \end{enumerate}

%  \textcolor{red}{
\vskip.2cm
  Finally, the trade-network possesses a very important feature: the
  incentive of nodes to optimize their connectivity is measurable.
  To deepen our understanding of inferring structure from incomplete
  network information in the context of Economics, future research
  should be pursued in two main directions; new data and simulation:
  \begin{enumerate}
  \item
  We wish to locate industries that are more vulnerable to
  shocks. If link removal harms any industry that happens to hold
  most of the highly connected firms, under a dissortative regime
  the network could probably break up completely.
  It is worthwhile locating data sets of trade credit transactions
  in other countries. In the US, trade-bills can be traded in the
  market. So there are other factors that may affect the structure
  of that trade-network. We speculate that the US industrial trade
  network also appears dissortative in respect of degree correlations.
  \item
  Simulating an assortative network is relatively easy using priors
  like top-cap on the firm size (see \cite{Huisman:2009fk}). Simulating
  the conditions that lead a popularity-based network to visually appear
  dissortative may prove useful at this point.
  \end{enumerate}
%  }

%  find ways to
%  locate neighborhoods where data is more likely to be missing.

\section*{Acknowledgment}
  We acknowledge the support of the Institute for New Economic
  Thinking (INET), inaugural grant number IN01100017.


\bibliographystyle{IEEEtran}
%\bibliography{IEEEabrv}
\bibliography{biblio_main}

\begin{appendix}
  \renewcommand{\arraystretch}{1.3}
  \begin{table}[htb]
    \centering
  \caption{node attribute names and their meaning}
  \begin{tabular}{|c|l|c|}
    \hline
    \textbf{name} & \textbf{description} & \textbf{source}\\
    \hline
    NACE & 1 or 2-digit industrial classification of a firm \cite{eurostat:fk} & BS\\
    RATING & credit-rating score $\{1,2 ,\dots ,9,99\}$. & \\
           &  $99$ is unclassified & BS \\
    Size & firm net sales [EUR] & BS\\
    FC & financial costs (interest paid) [EUR] & BS \\
    TBL & total bank loans [EUR] & BS \\
    $\Kin$ & number of known buyers & TC\\
    $P_{ji}$ & a payment from buyer $j$ to $i$ & TC\\
    \a & information exposure. See equation (\ref{eq:a}) & TC,BS\\
    \hline
  \end{tabular}
  \end{table}

  Note: The NACE (European) industrial classification scheme is a
  hierarchical numbering system. The leftmost digit is the major industry
  code. Further resolution of sub-classifications can be achieved
  by adding less significant digits, up to 4 digits in the case of
  NACE.

\end{appendix}

\end{document}